\documentclass[aps,prl,reprint,longbibliography,nofootinbib,superscriptaddress,floatfix]{revtex4-2}

\usepackage{appendix}
\usepackage{verbatim}
\usepackage[T1]{fontenc}
\usepackage{mathbbol}
\usepackage[dvipsnames]{xcolor}
\usepackage{orcidlink}
\usepackage[english]{babel}
\usepackage{lipsum}
\usepackage{physics}
\usepackage{dcolumn}
\usepackage{tensor}
\usepackage{comment}
\usepackage{graphicx,color,overpic,mathtools}
\usepackage{amsthm,amsmath,amssymb,mathrsfs}
\usepackage{braket,bm,bbm,setspace}
\usepackage{float}
\usepackage{xargs}
\definecolor{myred}{RGB}{179, 27, 27}
\definecolor{teal}{RGB}{0, 128, 128}
\definecolor{myblue}{RGB}{71, 81, 200}
\usepackage{hyperref}
\hypersetup{
	colorlinks=true,
	linkcolor=myblue, 
	citecolor=myblue, 
	urlcolor=myblue  
}
\hypersetup{hypertexnames=false}

\begin{document}
\title{Black-hole ringdown at an exceptional point}
\author{Kyriakos Destounis}
\affiliation{CENTRA, Departamento de F\'isica, Instituto Superior T\'ecnico -- IST, Universidade de Lisboa -- UL, Avenida Rovisco Pais 1, 1049-001 Lisboa, Portugal}
\affiliation{School of Applied Mathematical and Physical Sciences -- SEMFE, Physics Department, National Technical University of Athens -- Polytechnic Campus, Zografou 15772, Athens, Greece}
\begin{abstract}
Black holes ring in damped oscillations. At an exceptional point, photon-sphere and near-horizon modes merge, and their response can acquire a linear time factor. We compute its amplitude from constrained Einstein–Maxwell data for a rotating, charged black hole in de Sitter space. With seven competing modes fixed, the late-time curvature error is 0.00383\% for the pair and 0.01401\% for a refitted exponential. We suppress the linear term by tuning the initial disturbance at fixed geometry. Hence, near coalescence, changing only the initial disturbance can shift the frequency inferred from a short ringdown fit.
\end{abstract}

\maketitle

\textit{Introduction.---}Perturbed black holes create gravitational waves that can escape to infinity or fall through their event horizons. During ringdown, much of the signal can be described by exponentially-damped oscillations called quasinormal modes (QNMs) \cite{Kokkotas:1999bd,Berti:2009kk}, though QNMs alone do not determine the whole ringdown waveform, due to late-time tails \cite{Price:1971fb,Price:1972,Gundlach:1993tp,Gundlach:1993tn}. Their frequencies and decay rates characterize the black hole. The strength of each oscillation depends on the disturbance. Dissipation through the boundaries makes the wave problem non-Hermitian (non-self-adjoint), which can give rise to spectral instabilities \cite{Jaramillo:2020tuu,Cheung:2021bol,Jaramillo:2021tmt,Destounis:2021lum,Jaramillo:2022kuv,Boyanov:2022ark,Boyanov:2023qqf,Destounis:2023nmb,Cownden:2023dam,Arean:2023ejh,Cao:2025afs,Courty:2023rxk,Destounis:2025dck,Yang:2024vor,Cardoso:2024mrw,Destounis2026ExceptionalLines,Destounis:2026ays,Zhang:2026cjq,Berti:2022xfj,May:2026qpf}, and exceptional points (EPs). At an EP, two modes coalesce in both frequency and eigenfunctions. The merged pair can then contribute $(A+Bt)e^{-i\omega_{\rm EP}t}$, with a term linear in time multiplying a decaying oscillation. The time factor is familiar from critical damping and also appears in oscillating open systems such as microwave cavities \cite{Heiss2012,MaassenYoung2001Jordan,DietzEtAl2007Microwave}. The amplitude of the linear-in-time term depends on how the black hole is disturbed and on which quantity is measured \cite{CartariusMoiseyev2011Survival,HashemiEtAl2022EPResponse}.

Black-hole calculations have identified resonant excitation near avoided crossings \cite{Motohashi2025Resonant,TakahashiEtAl2025Coupled}, and located EPs and mode exchange in massive-scalar perturbations of Kerr \cite{Cavalcante:2024swt}. EPs also occur in systems with coupled perturbations \cite{Destounis2026ExceptionalLines,JenksEtAl2026Coupled}. In Ref. \cite{YangBertiFranchini2025}, the authors connected the exchange of modes around an EP to a linear-in-time waveform term and showed how it can bias frequency estimates in a model. Closely related studies have examined waveforms in a perturbed Schwarzschild potential, gravitational ringdown near avoided crossings in Kerr, and the scalar response near and at coalescence in the Nariai limit \cite{MacedoEtAl2026EPRingdown,KubotaMotohashi2026Ringdown,NakamotoOshita2026Nariai}. Scalar evolutions on hairy black holes have tested the polynomial response at a numerically tuned EP and shown how competing signals affect inference \cite{ChengEtAl2026HairyEP,Zhang:2026uoo}. Calculations of excitation factors in Kerr--de Sitter and Kerr--Newman characterize the resonances of the background \cite{Oshita:2025ibu,RossiEtAl2026KdSExcitation,ZhaoZi2026Excitation}.

Here we ask how strongly a specified disturbance excites the merged pair when gravity and electromagnetism evolve together. We compute $A$ and $B$ from initial data that satisfy the linearized Einstein--Maxwell constraints and predict the resulting curvature with seven competing modes included. One resonance belongs to the family associated with unstable photon orbits. The other is associated with the nearly extremal event horizon. We follow their contribution across coalescence and test it against the evolved curvature. At fixed black-hole parameters, changing the relative orientation and strength of two initial disturbances can suppress the coefficient $B$.

\begin{figure*}[t]
	\centering
	\includegraphics[width=\textwidth]{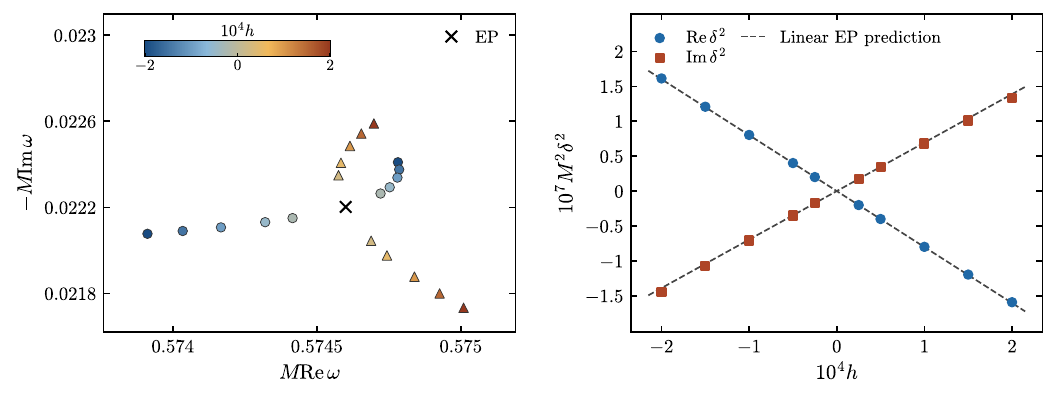}
	\caption{Splitting of the exceptional pair across neighboring stationary geometries. Left: both poles $\omega_\pm$ at each nonzero spin detuning $h=a-a_{\rm EP}$ at fixed $\Lambda$ and $\epsilon=1-r_-/r_+$. Color gives the signed detuning, with circles for $h<0$ and triangles for $h>0$. The black cross marks the exceptional point. Right: real and imaginary parts of the squared half-gap $\delta^2=[(\omega_+-\omega_-)/2]^2$ against signed $h$. Dashed lines show the leading-order prediction from the matching-function derivatives at the EP.}
	\label{fig:unfolding}
\end{figure*}

\textit{Locating the exceptional point.---}We study a rotating, charged black hole with a positive cosmological constant, the Kerr--Newman--de Sitter solution. We use geometrized units $G=c=1$, set the mass $M=1$ for numerical values, and use the Fourier decomposition $e^{-i\omega t+im\phi}$ in Boyer--Lindquist coordinates. The parameters $a$, $Q$, and $\Lambda$ specify rotation, charge, and the cosmological constant. The Cauchy, event, and cosmological horizons lie at $r=r_-$, $r=r_+$, and $r=r_c$, respectively. We measure proximity to Cauchy--event horizon merger by $\epsilon=1-r_-/r_+$. At the $m=2$ EP, $(a_{\rm EP},Q_{\rm EP},\Lambda)\simeq(0.31482907,0.94933058,0.001)$, $\epsilon\simeq0.00912130$, and
\begin{equation}
 \omega_{\rm EP}\simeq0.57460039290-0.02220194797i.
 \label{eq:ep}
\end{equation}

We use the separation of the coupled radiative fields from Ref. \cite{Zhao2026KNAdS}, with related formulations in Refs.~\cite{Hintz2026KN,DestounisCardosoHintz2026,Yoshida2026KNdS,WellerEtAl2026Parity}. On the selected angular branch $\lambda(\omega)$, let $E$ and $C$ be the radial solutions for waves entering the event horizon and leaving through the cosmological horizon. At a QNM frequency these solutions match, giving a zero of $D(\omega)=\det[E(3;\omega),C(3;\omega)]$ in the normalization $E_2(3)=C_2(3)=1$. At the EP, $D=D_\omega=0$, $D_{\omega\omega}\ne0$, and the radial boundary problem has one independent solution. A loop around the EP in spin and horizon separation exchanges the two frequencies. Following the modes away from the encounter connects one to a gravitational-led photon-sphere (PS) mode and the other to a near-horizon (NH) mode, as in the families studied in Kerr--Newman \cite{DiasEtAl2022Repulsion,DiasGodazgarSantos2022Extended}. Appendix~\ref{app:geometry} gives the geometry, normalizations, and mode-tracking checks.

The positive $\Lambda$ shifts the extremal spin--charge bound. At the central EP, $a_{\rm EP}^2+Q_{\rm EP}^2$ exceeds the flat-space horizon bound $M^2$. Retuning spin and charge as $\Lambda$ decreases gives a subextremal EP at $\Lambda=0$ with the outgoing condition at infinity (Appendix~\ref{app:flat-continuation}). The waveform calculations below are at $\Lambda M^2=0.001$. For a separate check of the spectrum, the flat-limit solver agrees with published coupled partial-differential-equation (PDE) results at three Kerr--Newman backgrounds to within $1.5\times10^{-11}$ in $M\omega$ \cite{DiasGodazgarSantos2015,DestounisCardosoHintz2026}. Appendix~\ref{app:flat-benchmarks} includes direct coupled-equation comparisons and a positive-$\Lambda$ convergence test.

\textit{From an initial disturbance to curvature.---}The Einstein--Maxwell constraints relate the electric and magnetic data to the spatial geometry and its initial rate of change. We start with a divergence-free electric seed and solve the linearized constraints on a maximal stationary slice \cite{IsenbergMaxwell2021}. The inner, central, and outer seeds differ in the location and width of their radial Gaussian envelopes. Appendix~\ref{app:radiative-map} gives the seed formula, the three profiles, and the metric and extrinsic-curvature corrections. We form real data by adding the $m=2$ sector to its conjugate. Axisymmetry allows us to evolve and extract the $m=+2$ component separately.

Our observable is the complex $m=+2$ azimuthal Fourier coefficient $\mathcal O=(\delta\Psi_0)_{m=+2}$ of the radiative curvature, extracted at $r=3M$, $\theta=\pi/2$.. We use a fixed principal frame. In that frame the background $\Psi_0$ and $\Psi_1$ vanish, so this curvature perturbation is invariant under infinitesimal coordinate and tetrad gauge changes. The conversion to a local Carter observer's frame and proper time preserves the pole order (see Appendix~\ref{app:radiative-map}).

\begin{figure*}[t]
	\centering
	\includegraphics[width=\textwidth]{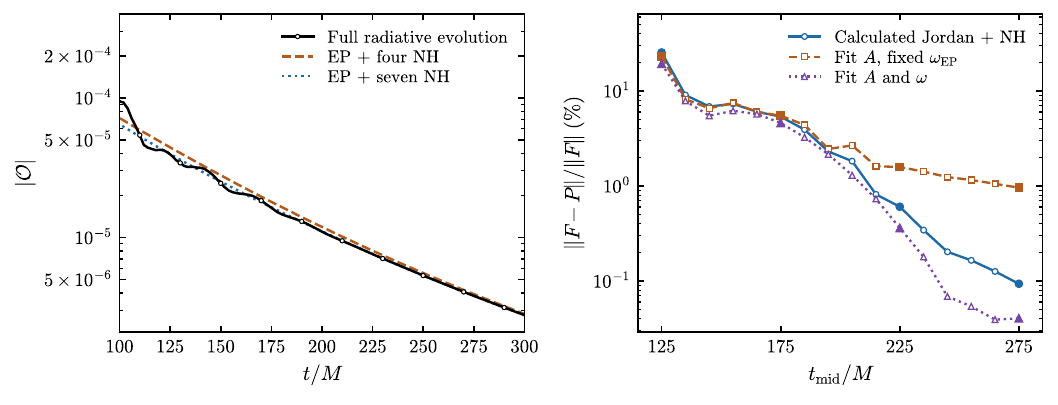}
	\caption{Pole-sum and exponential comparisons for the $m=+2$ azimuthal Fourier coefficient $\mathcal O$ of $\delta\Psi_0$. Left: $|\mathcal O|$ from the central-source evolution and from pole sums with four or seven NH terms. Circles mark selected evolution samples. Right: relative Euclidean norm discrepancies of the complex samples, in percent, for $50M$ windows shifted by $10M$ and plotted at their midpoints. Filled markers identify the four windows in Table~\ref{tab:central-fits}. All right-panel comparisons retain seven fixed NH terms. The Jordan curve uses the calculated pair contribution $(A+Bt)e^{-i\omega_{\rm EP}t}$. The other curves replace it by a damped exponential fitted as indicated.}
	\label{fig:waveform}
\end{figure*}

Initial data enter the frequency-domain calculation as a forcing term. For the radiative state $q(t)$ and operator $\mathcal{Q}(\omega)=\mathcal{Q}_0+\omega\mathcal{Q}_\omega$, the one-sided transform gives
\begin{equation}
 \mathcal{Q}(\omega)\widetilde q=i\mathcal{Q}_\omega q(0),\qquad
 \widetilde q=\int_0^\infty e^{i\omega t}q(t)\,\mathrm{d}t.
 \label{eq:forcing}
\end{equation}
Here, $\mathcal{Q}_\omega=\partial_\omega\mathcal{Q}$. This relation lets us calculate how much of each pole enters $\mathcal O$. We also evolve the constrained initial state on a radial--angular grid and compare with the pole prediction. The time-domain and pole calculations share the radiative equations and initial-data map (see Appendices~\ref{app:radiative-map} and \ref{app:evolution-detuning}).

\textit{How the two modes merge.---}At the EP, the part of the retarded curvature transform from the selected angular branch has the expansion $\widetilde{\mathcal O}_{\rm br}=C_{-2}/(\omega-\omega_{\rm EP})^2+C_{-1}/(\omega-\omega_{\rm EP})+O(1)$. The two Laurent coefficients set the amplitudes in the time signal $(A+Bt)e^{-i\omega_{\rm EP}t}$ through $A=-iC_{-1}$ and $B=-C_{-2}$. We compute them by projecting the initial data onto the response at the pole. The stationary initial slice meets the event-horizon bifurcation sphere, so we evaluate the source integrals first where they converge in the upper half of the frequency plane and then continue them to the pole (see Appendix~\ref{app:green-functional}). For the central disturbance we obtain
\begin{equation}
 C_{-2}=(-7.18045-8.80768i)\times10^{-7},
 \label{eq:c2}
\end{equation}
with a relative change of $9.4\times10^{-7}$ under source refinement. The same constraint solution gives $A\simeq(-5.230+3.186i)\times10^{-4}$.

A small shift in spin splits the double pole back into two ordinary poles. We vary $a=a_{\rm EP}+h$ and adjust $Q$ to hold $\epsilon$ fixed, with $\Lambda$ also fixed. For the waveform comparisons we use $h=\pm5\times10^{-5},\pm2\times10^{-4}$. The spectral and residue-moment figures also sample $h=\pm2.5\times10^{-5},\pm10^{-4},\pm1.5\times10^{-4}$. We solve the constraints afresh for the residue and waveform calculations, preserving the central seed prescription and extraction point. Let $\bar\omega=(\omega_++\omega_-)/2$ be the midpoint frequency, $\delta=(\omega_+-\omega_-)/2$ the half-gap, and $R_\pm$ the residues of $\widetilde{\mathcal O}_{\rm br}$ at the two poles. The combinations $C_{-1}(h)=R_++R_-$ and $C_{-2}(h)=\delta(R_+-R_-)$ give the pair's contribution as
\begin{equation}
 \mathcal O_{\rm pair}(t)=e^{-i\bar\omega t}
 \left[-iC_{-1}(h)\cos(\delta t)-C_{-2}(h)\frac{\sin(\delta t)}{\delta}\right].
 \label{eq:pair}
\end{equation}
As the poles merge, the sine quotient tends to $t$. Thus, the pair has a finite limit even as its individual residues grow. A related contour construction gives the sourced coefficients from the resolvent of the pair \cite{MorikawaEtAl2026RieszLaurent}. The double zero also predicts a square-root splitting. In terms of the squared half-gap, this reads $\delta^2=K_{\rm EP}h+O(h^2)$, with $K_{\rm EP}=-2D_a/D_{\omega\omega}$ evaluated at the EP. Here $D_a$ differentiates the matching determinant with respect to $a$ at fixed $\omega$, $\Lambda$, $\epsilon$ and $M$, with $Q$ adjusted along this path and the selected angular eigenvalue substituted. The computed complex $\delta^2$ differs from this independent linear prediction by less than $2.8\%$ of $|K_{\rm EP}h|$ at all ten detunings as demonstrated in Fig.~\ref{fig:unfolding}. The log--log slopes of $|\omega_+-\omega_-|$ against $|h|$, fitted to the five detunings on each side, are $0.5043$ for $h<0$ and $0.4957$ for $h>0$. Both residue combinations approach their EP values approximately linearly in $|h|$ (see Fig.~\ref{fig:residue-moments} in Appendix \ref{app:green-functional}). This tests the limit of the sourced response as well as the splitting of the frequencies.

\begin{figure*}[t]
	\centering
	\includegraphics[width=\textwidth]{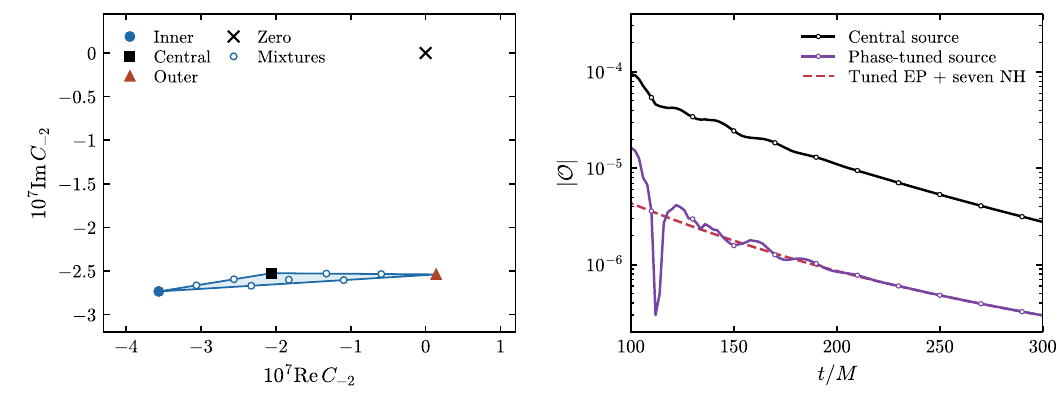}
	\caption{Source dependence at fixed geometry. Left: double-pole coefficients $C_{-2}$, which set the linear-term amplitude $B=-C_{-2}$, for the three seeds scaled to unit electric-seed norm. Open circles are coefficients for nonnegative mixtures of these rescaled seeds with weights summing to one. Their convex hull excludes zero. Right: $|\mathcal O|$, the modulus of the $m=+2$ azimuthal Fourier coefficient of $\delta\Psi_0$, for the central source, the tuned combination of separately evolved central and inner data, and the tuned pole prediction. Circles mark selected evolution samples. The curves retain the stated seed amplitudes and relative weight.}
	\label{fig:sources}
\end{figure*}

\textit{Testing the ringdown.---}The merging modes are only part of the signal. Seven other NH modes decay slowly enough to compete with the pair in our waveform (Table~\ref{tab:nh-poles}). For source profile $j$ we predict
\begin{equation}
 P_j(t)=(A_j+B_jt)e^{-i\omega_{\rm EP}t}
 -i\sum_{n\in\{0,1,2,3,5,6,7\}}R_{nj}e^{-i\omega_nt}.
 \label{eq:prediction}
\end{equation}
The complex weights $A_j,B_j,R_{nj}$ all come from the same constraint solution used in the evolution. Here $R_{nj}$ is the residue at the NH pole $\omega_n$ for profile $j$, and $n=4$ labels the NH branch that enters the merged pair. We compare the complex curvature samples $F_j$ with $P_j$ evaluated at the same times, using the relative Euclidean norm $\|F_j-P_j\|/\|F_j\|$. The two fundamental de Sitter modes, associated with the cosmological horizon, contribute less than $3\times10^{-10}$ of the evolved norm on $250\leq t/M\leq300$ for the tested profiles and tuned mixture (see Appendix~\ref{app:de-sitter}).

For the central disturbance on $250\leq t/M\leq300$, the merged pair alone differs from the evolved signal by $67.3\%$, and the seven NH modes alone by $40.2\%$. Adding the pair to four NH modes reduces the difference to $3.50\%$. Including all seven gives $0.0935\%$ (Fig.~\ref{fig:waveform}). For the inner and outer profiles the corresponding errors are $0.0746\%$ and $0.1131\%$. Earlier, on $100\leq t/M\leq150$, the central error is $25.4\%$. This time dependence delineates the interval where the calculated pole sum describes the evolution accurately. On the four detuned backgrounds with evolutions, the split pair and updated NH contributions give $0.0929$--$0.0941\%$ on the late window. The spectral and residue tests above track the approach to coalescence, while the time-domain changes also reflect shifts in the midpoint and the source coupling (Appendix~\ref{app:evolution-detuning} and Table~\ref{tab:detuned-waveforms}).

Can a shifted ordinary frequency mimic the linear term? On $250\leq t/M\leq300$, deleting $Bt$ with every other term fixed raises the central error to $18.8\%$. Refitting only the amplitude gives $0.961\%$. Refitting both amplitude and complex frequency lowers the error to $0.0405\%$, below the $0.0935\%$ of the calculated pole sum on that window. The pair's instantaneous complex frequency $\omega_{\rm EP}+iB/(A+Bt)$ explains how a constant shift can approximate its time dependence over a short interval (see Appendix~\ref{app:evolution-detuning}).

The longer evolution gives a different comparison. On $350\leq t/M\leq500$, the calculated pole sum differs from the evolved curvature by $0.00383\%$, replacing the pair by a single exponential refitted on that interval gives $0.01401\%$, and carrying the earlier fit forward gives $0.26207\%$. Each comparison keeps the same seven NH contributions. The ranking holds on all three grids and for a second radiative invariant. Between the two finest grids the curvature waveform changes by $0.00052\%$. Its root-mean-square magnitude on the late interval is $4.78\times10^{-5}$ of the pulse peak (Appendix~\ref{app:extended-waveforms}, Table~\ref{tab:extended-ivp}).

\textit{Controlling the excitation.---}Changing the initial perturbation leaves the EP frequency fixed but can change how strongly its linear term is excited. For each of our three seeds, the normalized coefficient satisfies $-\operatorname{Im}(C_{-2}/N_e)>2.5\times10^{-7}$, where $N_e$ is the electric-seed norm (Fig.~\ref{fig:sources}). Nonnegative mixtures share this sign. Their predicted and evolved curvature agree to within $0.116\%$ on $250\leq t/M\leq300$, using the same normalization for the two signals (Appendix~\ref{app:green-functional}).

The double-pole coefficient vanishes if the source has zero overlap with the left EP mode or if the observable has zero overlap with the right EP mode \cite{HashemiEtAl2022EPResponse,MorikawaEtAl2026RieszLaurent}. Here an azimuthal rotation supplies the relative phase needed for cancellation. In the $m=+2$ sector, we add the central data to $w_{\rm mix}$ times the inner data, with $w_{\rm mix}=-B_{\rm central}/B_{\rm inner}$. The calculated weight rescales the inner seed by $1.1564$ and shifts its azimuthal argument by $-83.34^\circ$. Adding the conjugate sector gives real constrained data. On $250\leq t/M\leq300$, the combined curvature norm is $0.09661$ of the central norm and the pole sum reproduces the combined evolution to $0.169\%$. The change in total curvature includes the altered amplitudes of the ordinary poles.

\textit{Conclusions and discussion.---}A black hole at an exceptional point can ring with a linear time factor multiplying the damped oscillation at the merged frequency. The spectrum fixes that frequency and permits the linear-in-time term. The initial perturbation and the chosen curvature observable determine the amplitude of the linear-in-time term. Here those inputs predict both a nonzero coefficient and its suppression through a relative azimuthal shift and rescaling of the initial data.

Close to coalescence, the two frequencies and their separate residues can be difficult to interpret as independent oscillations. The midpoint, squared half-gap, and combined residues in Eq.~\eqref{eq:pair} are invariant under exchange of the two poles and describe their interference through the merger. The waveform comparisons show why the distinction matters. A shifted exponential can absorb the time dependence on a short interval. Over a longer interval the calculated polynomial term gives better agreement in this curvature channel when the other resonances are included.

For black-hole spectroscopy, this motivates testing the response of a resonant pair across several time windows with the competing modes included. Applying this test to a merger would require the remnant perturbation and a prediction for the corresponding radiative signal. The calculated finite-radius signal provides a benchmark for connecting such data to the excitation of the pair.
\\
\begin{acknowledgments}
\textit{Acknowledgments.---}K.D. acknowledges financial support provided by FCT--Funda\c{c}\~ao para a Ci\^encia e a Tecnologia, I.P., under the Scientific Employment Stimulus -- Individual Call -- Grant No. 2023.07417.CEECIND/CP2830/CT0008. K.D. would also like to thank the Funda\c{c}\~ao para a Ci\^encia e Tecnologia (FCT), Portugal, for the financial support to the Center for Astrophysics and Gravitation (CENTRA/IST/ULisboa) through grant No. UID/PRR/00099/2025 and grant No. UID/00099/2025. This project has received funding from the European Union's Horizon MSCA-2022 research and innovation programme ``Einstein Waves'' under grant agreement No. 101131233.
OpenAI ChatGPT and Codex (GPT-6 family) assisted with numerical exploration, plotting, literature scans, and manuscript auditing under author-specified research targets. The numerical comparisons and refinements used to check the calculations are described in the Letter and Supplemental Material.
\end{acknowledgments}

\paragraph*{Data availability.} The numerical data, solver and plotting scripts, and validation records are available upon reasonable request.
\\
\appendix
\raggedbottom
\setcounter{secnumdepth}{1}

\begin{center}
{\Large\bf Supplemental material}
\end{center}

\section{Geometry, boundary problem and mode assignment}\label{app:geometry}
We use $G=c=M=1$, $\Sigma=r^2+a^2\cos^2\theta$, $T^2=1+\Lambda a^2\cos^2\theta/3$, $\Xi=1+\Lambda a^2/3$ and $\Delta=(r^2+a^2)(1-\Lambda r^2/3)-2Mr+Q^2$. The Boyer--Lindquist metric is
\begin{equation}
 \begin{aligned}
 ds^2={}&-\frac{\Delta}{\Sigma}\left(\mathrm{d}t-\frac{a\sin^2\theta}{\Xi}\mathrm{d}\phi\right)^2
 +\frac{\Sigma}{\Delta}\mathrm{d}r^2+\frac{\Sigma}{T^2}\mathrm{d}\theta^2\\
 &+\frac{T^2\sin^2\theta}{\Sigma}\left(a\mathrm{d}t-\frac{r^2+a^2}{\Xi}\mathrm{d}\phi\right)^2.
 \end{aligned}
\end{equation}
Here $\phi$ has period $2\pi$ and the Fourier convention is $e^{-i\omega t+im\phi}$. The background Maxwell potential is $\mathcal A_{\rm EM}=-Qr(\mathrm{d}t-a\sin^2\theta\,\mathrm{d}\phi/\Xi)/\Sigma$, with $F_{ab}=(\mathrm{d}\mathcal A_{\rm EM})_{ab}$. Our frequency equals Zhao's and $\Xi$ times Yoshida's frequency \cite{Zhao2026KNAdS,Yoshida2026KNdS}. The positive surface gravities are $\kappa_+=\Delta'(r_+)/[2(r_+^2+a^2)]$ and $\kappa_c=-\Delta'(r_c)/[2(r_c^2+a^2)]$. For $\nu\in\{+,c\}$, the angular velocity of the horizon at $r_\nu$ is $\Omega_\nu=a\Xi/(r_\nu^2+a^2)$. We approach the cold extremal limit $r_-\to r_+$.

We implement the de Sitter specialization of Zhao's separated two-component radial and angular system between the event and cosmological horizons, using Yoshida's conventions for comparison \cite{Zhao2026KNAdS,Yoshida2026KNdS}, in the positive radiative representation $\varsigma=+1$ and $m=2$. This representation contains coupled gravitational and electromagnetic degrees of freedom, rather than a test spin field. With Pauli matrices $\sigma_k$, these systems have the form
\begin{equation}
 \begin{aligned}
 \sqrt\Delta\,u'&=[R_0\sigma_3+R_1\sigma_1+(h_r+\lambda)\sigma_2]u,\\
 T\chi'&=[-\Theta_1\sigma_1-\Theta_0\sigma_3+i(\lambda-h_\theta)\sigma_2]\chi.
 \end{aligned}
\end{equation}
Primes denote derivatives with respect to $r$ for $u$ and $\theta$ for $\chi$. The coefficient and algebraic-frame definitions are those of Eqs.~(9), (12) and (19)--(23) of Ref.~\cite{Zhao2026KNAdS} (arXiv v1), with $\Delta$ and $T$ as above. Let $s_\theta=\theta$ at the north axis and $s_\theta=\pi-\theta$ at the south axis, with the corresponding exponent parameters $\alpha_N=\varsigma m+3/2$ and $\alpha_S=3/2-\varsigma m$. The regular solution is $s_\theta^{|\alpha|}(f_1(s_\theta^2),s_\theta f_2(s_\theta^2))^{\mathsf T}$ for $\alpha>0$ and $s_\theta^{|\alpha|}(s_\theta f_1(s_\theta^2),f_2(s_\theta^2))^{\mathsf T}$ for $\alpha<0$.

Put $H=\exp(i\int K_r\,\mathrm{d}r/\Delta)$, $K_r=a\Xi m-(r^2+a^2)\omega$, and $\zeta_+=+1$, $\zeta_c=-1$. The ingoing event-horizon and outgoing cosmological-horizon branches have
\begin{equation}
 \begin{aligned}
 u_1&\in H^{\zeta_\nu}\Delta^{1/4-\zeta_\nu\varsigma/2}\mathcal A_\nu,\\
 u_2&\in H^{\zeta_\nu}\Delta^{1/4-\zeta_\nu\varsigma}\mathcal A_\nu,
 \end{aligned}
\end{equation}
where $\mathcal A_\nu$ denotes regular Frobenius amplitudes at horizon $r_\nu$. The half-power component offset is required. Boundedness of an arbitrarily rescaled field does not select the physical branch.

Let $D_{\rm ang}(\omega,\lambda)$ be the matching determinant of the north- and south-regular angular solutions. Its simple zero in $\lambda$ defines the selected branch $\lambda(\omega)$, with $\lambda_\omega=-D_{{\rm ang},\omega}/D_{{\rm ang},\lambda}$ and $\lambda_{\rm EP}\simeq-0.9207778996333+0.1336461138223i$. On this branch, let $F_{r_m}(\omega)=E_1(r_m)/E_2(r_m)-C_1(r_m)/C_2(r_m)$ at matching radius $r_m$. The radial coefficient matrix has zero trace, so its propagator has unit determinant. Consequently, $\det[E(r_m),C(r_m)]$ is independent of $r_m$ for fixed horizon normalizations. In a chart where $E_2(r_m)C_2(r_m)\ne0$, $F_{r_m}=\det[E(r_m),C(r_m)]/[E_2(r_m)C_2(r_m)]$. Changing $r_m$ within such a chart multiplies $F_{r_m}$ by a nonzero analytic factor. For the transfer calculation we set $r_m=3$ and normalize $E_2(3)=C_2(3)=1$, so $D=F_3$.

The central numerical locator is $(a_{\rm EP},Q_{\rm EP},\Lambda)\simeq(0.3148290718469,0.9493305753700,0.001)$, with $r_+\simeq1.0052859053$, $r_c\simeq53.7523632070$ and $\omega_{\rm EP}\simeq0.5746003929047-0.0222019479707i$. At the production angular settings, the 40- and 50-digit radial calculations at $r_m=2$ give $|F_2|<8.4\times10^{-13}$ and $|\partial_\omega F_2|<3.3\times10^{-10}$ across the central and two neighboring retuned $\Lambda$ slices. For the central point, changed angular settings at $r_m=2$ give $1.4\times10^{-12}$ and $1.3\times10^{-10}$, and changing to $r_m=3$ gives approximately $1.1\times10^{-11}$ and $1.1\times10^{-9}$. The angular inputs remain double precision. The radial matching matrix with unit-normalized columns has singular values near $\sqrt{2}$ and in the range $4\times10^{-14}$--$1.3\times10^{-13}$, consistent with a one-dimensional kernel. In the $r_m=3$ chart, an independent scattering calibration gives $\tfrac12\partial_\omega^2F_3\simeq 2.41206\times10^7-2.00596\times10^7i$, reproduced by extrapolated frequency differences to $7.3\times10^{-7}$ relative. This coefficient depends on the chart.

The two-real-control unfolding in $(a,\epsilon)$, with $\epsilon=1-r_-/r_+$, has rank two and condition number approximately $4.74$ in the chosen coordinates. Corrected gaps give a log--log slope of $0.500256$ over the three smallest radii on the splitting ray. On a 24-step transverse loop of maximum control semiaxis $10^{-6}$, the two sheets exchange with a swap error of $1.3\times10^{-9}$ compared with initial gap $5.34\times10^{-5}$. Together these tests identify the numerical order-two EP. On the spin path $h=a-a_{\rm EP}$ at fixed $\epsilon$ and $\Lambda$, the leading squared half-gap is $K_{\rm EP}h$, with $K_{\rm EP}=-2(\partial_a F_3)_{\epsilon,\Lambda}/\partial_\omega^2F_3\simeq(-8.016+6.946i)\times10^{-4}$ in units $M=1$. The derivatives retain the angular eigenvalue dependence and are evaluated at the EP. This gives the dashed lines in Fig.~\ref{fig:unfolding}.

We checked the locator independently with multiprecision angular Chebyshev--Gauss collocation after factoring the regular axis weights. We solved $F_3=\partial_\omega F_3=0$ with 32 points per angular component and multiprecision radial propagation. The changes in $a,\epsilon,\operatorname{Re}\omega_{\rm EP},\operatorname{Im}\omega_{\rm EP}$ are below $2\times10^{-15}$ each. Evaluation at 36 points with changed radial starts, Taylor settings and Frobenius orders gives $|F_3|=4.16\times10^{-16}$ and $|\partial_\omega F_3|=4.22\times10^{-13}$. The second derivative is $(4.8241256285-4.0119233180i)\times10^7$ in this chart. The waveform calculation uses the original working values. The printed coordinates are rounded.

At the original working point the sampled radial determinant winds twice on each of two frequency circles of radii $2\times10^{-4}$ and $4\times10^{-4}$, with maximum adjacent phase increments below $0.90$ rad. At the center and 32 samples of the larger circle, angular collocation finds a gap of at least $1.27$ to the nearest other angular eigenvalue. These checks concern the tracked angular eigenvalue and its radial boundary problem.

For the PS connection, an ordinary gravitational-led fundamental at $a=a_{\rm EP}$, $Q=0.94$, $\Lambda=0.001$ has $\omega=0.56810829244-0.04910627620i$. The finite-damping sheet near the EP is detoured through $a=a_{\rm EP}-10^{-4}$ and $\epsilon=0.0102$, restored to $a_{\rm EP}$, then continued to $a=0.29$, $Q=0.955$. Independent continuation from the ordinary fundamental reaches the same endpoint $\omega=0.558342867820-0.039154018595i$ within $1.7\times10^{-13}$. Forty accepted checkpoints and fourteen backward checks on the first leg give maximum backward defect $1.4\times10^{-11}$. The other sheet approaches local throat seed $n=4$ as $\epsilon$ decreases. The merged pair contains that sheet, leaving seven additional poles in Table~\ref{tab:nh-poles}. The ancestry labels refer to the specified continuation paths.

\begin{table}[tbp]
\caption{Seven additional full-equation poles at the central EP. Indices are local NH seeds. Residues for each source and observable are supplied in the numerical data.}
\label{tab:nh-poles}
\centering
\begin{tabular}{crr}
\hline\hline
$n$ & $\operatorname{Re}\omega_n$ & $-\operatorname{Im}\omega_n$\\\hline
0&0.57234276094448&0.00482968005396\\
1&0.57254174144658&0.00896956996591\\
2&0.57280360738548&0.01312423692731\\
3&0.57320479328589&0.01732467792284\\
5&0.57326095138268&0.02709370132253\\
6&0.57290847577786&0.03129561938966\\
7&0.57269509297030&0.03544833526288\\\hline\hline
\end{tabular}
\end{table}

\section{Continuation to the asymptotically flat spectrum}
\label{app:flat-continuation}
At $\Lambda=0.001$ the metric parameters have $a_{\rm EP}^2+Q_{\rm EP}^2=1.0003458858$, with $r_-\simeq0.99611639$ and $r_+\simeq1.00528591$. The same spin and charge at $\Lambda=0$ give a negative discriminant for $\Delta=r^2-2r+a^2+Q^2$. We follow the exceptional point by solving $F_2=\partial_\omega F_2=0$ for $(\operatorname{Re}\omega,\operatorname{Im}\omega,a,\epsilon)$ at successively smaller $\Lambda$, obtaining $Q$ from the horizon roots. The positive-$\Lambda$ solves use the future cosmological horizon condition and changed-setting radial and angular checks. The table shows that the retuned curve crosses the flat-space bound $a^2+Q^2=1$ between $\Lambda=3\times10^{-5}$ and $6\times10^{-5}$ while keeping a positive horizon gap.

At $\Lambda=0$ the event and Cauchy horizon radii satisfy $r_+=2/(2-\epsilon)$ and $r_-=2-r_+$. We impose outgoing behavior at infinity by factoring $e^{i\omega r}r^{2i\omega-3/2}$ from the positive-representation radial solution and starting its inverse-radius expansion along a complex ray where the outgoing exponential decays. The two radial solutions are matched at $r_m=2$, with the outgoing solution at infinity replacing the cosmological-horizon solution in $F_2$. An independent multiprecision angular and radial calculation reproduces the endpoint when varying angular collocation from 24 to 30 nodes per component, the ray angle and length from $(0.34,72)$ to $(0.49,94)$, and the inverse-radius order from 42 to 56. It gives $|F_2|=1.33\times10^{-18}$ and $|\partial_\omega F_2|=1.64\times10^{-13}$ at the endpoint. The second frequency derivative is approximately $(1.715+4.276i)\times10^6$ in this chart, and the real two-control unfolding in $(a,\epsilon)$ has singular values $0.00331$ and $0.000699$. These are spectral continuation tests. All physical-source and time-evolution results in the Letter use $\Lambda=0.001$.

\begin{table*}[tp]
\caption{Sampled continuation of the exceptional frequency $\omega_{\rm EP}(\Lambda)$ and its geometry in units $M=1$. The $\Lambda=0$ row uses the outgoing condition at infinity. The gap variable at this endpoint is $\epsilon\simeq0.00914713$.}
\centering
\small
\begin{tabular}{rrrrrr}\hline\hline
$\Lambda$ & $a$ & $Q$ & $\operatorname{Re}\omega_{\rm EP}$ & $-\operatorname{Im}\omega_{\rm EP}$ & $10^5(a^2+Q^2-1)$\\\hline
$10^{-3}$&0.31482907&0.94933058&0.57460039&0.02220195&$34.58858$\\
$6\times10^{-4}$&0.31483779&0.94925030&0.57490354&0.02224906&$19.89706$\\
$3\times10^{-4}$&0.31484429&0.94919016&0.57513069&0.02228439&$8.88865$\\
$10^{-4}$&0.31484862&0.94915009&0.57528202&0.02230795&$1.55457$\\
$6\times10^{-5}$&0.31484948&0.94914208&0.57531228&0.02231265&$0.08822$\\
$3\times10^{-5}$&0.31485013&0.94913607&0.57533497&0.02231619&$-1.01144$\\
$0$&0.31485078&0.94913006&0.57535766&0.02231972&$-2.11101$\\\hline\hline
\end{tabular}
\end{table*}

\section{Flat-spectrum benchmarks against coupled PDEs}
\label{app:flat-benchmarks}
We test the separated spectrum at $\Lambda=0$ against three gravitational-led $\ell=m=2$ fundamentals from the author-supplied coupled-PDE table of Ref.~\cite{DiasGodazgarSantos2015}, archived on Berti's Ringdown data page. The source table uses $a/M=Q/M$ and lists $\operatorname{Im}(M\omega)$ before $\operatorname{Re}(M\omega)$. At each fixed geometry we track the angular eigenvalue from small rotation and solve the radial condition starting at a displaced complex frequency. Changing angular, horizon and outgoing-ray settings shifts the results by at most $1.4\times10^{-14}$ in $M\omega$. The smallest horizon gap of the three cases is $\epsilon\simeq0.03028$. The quoted differences measure agreement with the archived entries, whose accuracy is reported through ten decimal places in Ref.~\cite{DiasGodazgarSantos2015}.

\begin{table*}[tp]
\caption{Separated frequencies and absolute discrepancies from the coupled-PDE data of Ref.~\cite{DiasGodazgarSantos2015} at fixed Kerr--Newman backgrounds, with $\Delta\omega=\omega_{\rm sep}-\omega_{\rm PDE}$.}
\centering
\begin{tabular}{rrrr}\hline\hline
$a/M=Q/M$ & $\operatorname{Re}(M\omega)$ & $-\operatorname{Im}(M\omega)$ & $|M\Delta\omega|$\\\hline
0.296494222652&0.424021969277&0.087877276374&$1.36\times10^{-13}$\\
0.623623819067&0.591282979215&0.075076337806&$3.10\times10^{-14}$\\
0.707023201160&0.925309328970&0.004985017224&$1.45\times10^{-11}$\\\hline\hline
\end{tabular}
\end{table*}

We also solve the two coupled master PDEs of Ref.~\cite{DiasGodazgarSantos2015} directly at the first background, using the operators of its Supplemental Material. We factor the ingoing horizon, outgoing infinity and regular axis weights and collocate the two remaining fields on a tensor-product Chebyshev grid in radius and angle. With $44\times12$ radial--angular points, changing the complex-ray angle from $0.45$ to $0.72$ radians changes $M\omega$ by $4.7\times10^{-12}$. The resulting frequencies agree with the separated solution to within $10^{-11}$. The maximum off-grid residual divided by the sum of the magnitudes of the differential terms is below $10^{-11}$ at both ray angles for this grid.

For an additional two-root test we take the reference geometry of Eq.~(43) of Ref.~\cite{Destounis2026ExceptionalLines}, $(a/M,Q/M)=(0.3041149160474578,0.9523576757342892)$. The $90\times12$ coupled-PDE solves give $M\omega_1\simeq0.56519688405-0.02743179660i$ and $M\omega_2\simeq0.56364400681-0.02954173765i$, within $1.8\times10^{-10}$ of the separated values in absolute complex frequency. The pair gap is $|M(\omega_1-\omega_2)|\simeq2.62\times10^{-3}$. The eliminated PDE coefficients have poles in the complex radial plane at this charge. A three-domain calculation along a $1.38$-radian ray that avoids those coefficient poles reproduces each frequency within $2.4\times10^{-10}$ and agrees with the separated values to $7.5\times10^{-11}$. The numerical deposit gives the grid, contour and residual checks.

Holding the first benchmark's $a$ and $Q$ fixed at positive $\Lambda$, we obtain $|M(\omega(\Lambda)-\omega_{\mathrm{PDE}}(0))|=1.4325\times10^{-3},\,5.7211\times10^{-4},\,1.4292\times10^{-4}$ at $\Lambda M^2=10^{-3},\,4\times10^{-4},\,10^{-4}$, respectively. Changed-settings local frequency corrections are below $2\times10^{-9}$ across these points. This samples the approach to the external flat-spectrum benchmark with the cosmological horizon condition at positive $\Lambda$.

The coupled-PDE results of Refs.~\cite{DiasEtAl2022Repulsion,DiasGodazgarSantos2022Extended} establish PS and NH mode repulsion in flat Kerr--Newman. Their fixed-rotation repulsion plots display damping curves of the lowest branches. The complex-frequency reference used in the table is the author-supplied data from Ref.~\cite{DiasGodazgarSantos2015}. The spherical near-extremal zero-damped gravitoelectromagnetic spectrum was also studied in Ref.~\cite{ZimmermanMark2016}. These spectral comparisons test the flat-limit operator and boundary conditions. The sourced double pole and time-domain comparison of this work use the coupled radiative formulation at the positive-$\Lambda$ EP.

\section{De Sitter family and its contribution}
\label{app:de-sitter}
The cosmological-horizon family is distinct from the throat modes in the spherical charged problem \cite{DiasReallSantos2018RNdS,StuchlikZhidenko2026RNdS}. The small-hole estimate from pure de Sitter for the two fundamentals with $\ell=2$ is $-\operatorname{Im}\omega\simeq3\sqrt{\Lambda/3}$, and the central background has $\kappa_c=0.01757719$. At fixed $m=2$, both radiative sectors start at $\ell=2$ in the spherical limit. We track the gravitational-led and electromagnetic-led fundamentals from spherical master-equation anchors through a rotating intermediate point $(a,Q,\Lambda)=(0.2,0.3,0.001)$ to the central background. The ancestry names identify the limiting sectors of the coupled equations. Label these two poles by $\alpha\in\{\mathrm{g},\mathrm{e}\}$ for gravitational-led and electromagnetic-led ancestry. Their frequencies $\omega_\alpha^{\rm dS}$ are $0.00021039-0.05479035i$ and $0.00020993-0.05477273i$, respectively.

To check the lower damping range, we evaluated normalized radial determinants on a rectangular frequency contour centered horizontally at $m\Omega_c\simeq0.00021793$, with half-width $0.008$ and $0.007\leq-\operatorname{Im}\omega\leq0.045$. For both $\ell=2$ angular branches, the winding at 96 samples is $-1$ before removing the cosmological Frobenius normalization pole at $m\Omega_c-2i\kappa_c$ and zero after multiplication by $\omega-m\Omega_c+2i\kappa_c$. The count agrees between two endpoint offsets, Frobenius orders and matching radii. The largest phase step after regularization is below $0.10$ radians. The two branches with spherical $\ell=3$ ancestry also give zero regularized winding at 80 and 144 samples. This provides a numerical zero count on the tested branches in the specified rectangle, which covers damping near $2\kappa_c$.

For source $j$, the simple-pole residue is $R^{\rm dS}_{\alpha j}=N^{\rm dS}_{\alpha j}(\omega_\alpha^{\rm dS})/\partial_\omega D_\alpha(\omega_\alpha^{\rm dS})$, where $D_\alpha$ is the normalized radial matching determinant on the angular branch of pole $\alpha$. We compute the sourced numerator $N^{\rm dS}_{\alpha j}$ with the same initial data and the pole's own angular projection, using the endpoint continuations of Appendix~\ref{app:green-functional}. At these frequencies the cosmological overlap also needs the continuation. For the central data the curvature residues of the gravitational-led and electromagnetic-led poles are approximately $(-2.230+1.042i)\times10^{-10}$ and $(1.102+3.759i)\times10^{-11}$. Going from 128 to 160 radial source points changes these residues by $0.23\%$ and $0.17\%$. Under changed angular and radial settings the values of $\partial_\omega D_\alpha$ at the poles change by less than $6.1\times10^{-9}$ relative. The coherent contribution is $9.59\times10^{-9}$ of the central evolved curvature norm on $100\leq t/M\leq150$. The following late-window comparison uses the same complex samples and normalization as the Letter.

\begin{table}[tbp]
\caption{Relative curvature norms of the two de Sitter pole signals on $250\leq t/M\leq300$. The coherent column sums the complex signals before taking the norm. The last column sums their individual norms. Both divide by the corresponding evolved curvature norm.}
\centering
\begin{tabular}{lrr}\hline\hline
Source & Coherent sum & Sum of norms\\\hline
Inner   &$8.05\times10^{-12}$ &$8.64\times10^{-12}$\\
Central &$3.18\times10^{-11}$ &$3.56\times10^{-11}$\\
Outer   &$9.51\times10^{-11}$ &$1.33\times10^{-10}$\\
Tuned   &$2.44\times10^{-10}$ &$2.79\times10^{-10}$\\\hline\hline
\end{tabular}
\end{table}
\par\medskip

\section{Constrained data and the radiative map}
\label{app:radiative-map}
Let $n_\perp$ be the future unit normal, $h$ the spatial metric and $K_{ij}=-\mathcal L_{n_\perp}h_{ij}/2$. Let $D_i$ be the covariant derivative of $h$ and $|K|_h^2=K_{ij}K^{ij}$. We use $E_i=F_{ia}n_\perp^a$ and $F_{ij}=\varepsilon_{ijk}B^k$ with the spatial volume orientation $\mathrm{d}r\wedge\mathrm{d}x\wedge\mathrm{d}\phi>0$ in $x=\cos\theta$. The same volume form defines curls and cross products. With $\operatorname{tr}_hK=0$ and Einstein--Maxwell normalization $G_{ab}+\Lambda g_{ab}=2(F_{ac}F_b{}^c-g_{ab}F^2/4)$, the constraints read $R_h-|K|_h^2=2(E^2+B^2)+2\Lambda$, $D_jK^{ij}=2(E\times B)^i$, and $D_iE^i=D_iB^i=0$. The conformal construction \cite{IsenbergMaxwell2021} uses divergence-free seeds $e,b$, York operator $(LW)_{ij}=D_iW_j+D_jW_i-2h_{ij}D_kW^k/3$, and solves
\begin{equation}
 D_j(LW)^{ij}=2(e\times B+E\times b)^i,
\end{equation}
\begin{equation}
 \begin{aligned}
 &[-D^2+|K|_h^2+E^2+B^2-\Lambda]\varphi\\
 &\qquad=\tfrac14K_{ij}(LW)^{ij}+\tfrac12(E\cdot e+B\cdot b).
 \end{aligned}
\end{equation}
The perturbations are $\delta h_{ij}=4\varphi h_{ij}$, $\delta K_{ij}=(LW)_{ij}-2\varphi K_{ij}$, $\delta E^i=e^i-6\varphi E^i$ and $\delta B^i=b^i-6\varphi B^i$. This sets the independent conformal-metric and transverse-traceless seeds to zero. For $b=0$ the electric curl of $f(r)(1-x^2)e^{2i\phi}\mathrm{d}r$ has Fourier amplitudes $(0,2if(1-x^2),2xf)/\sqrt{\det h}$ in the oriented $(r,x,\phi)$ chart. The radial profiles are $f(r)=\exp[-(r-r_0)^2/w_0^2]$, with $(r_0,w_0)=(1.5,0.3),(2,0.5),(3,0.7)$. A real tensor perturbation is obtained by adding the conjugate Fourier sector, with a common freely chosen small overall amplitude.

The profile values at the event-horizon bifurcation sphere are approximately $0.066$, $0.019$ and $3.0\times10^{-4}$ for the inner, central and outer seeds. In a smooth transverse coordinate $s$ on the stationary slice, $r-r_+=O(s^2)$ and $\mathrm{d}r=O(s)\,\mathrm{d}s$. Since $\mathrm{d}(fY_{22}\,\mathrm{d}r)=f\,\mathrm{d}Y_{22}\wedge\mathrm{d}r$, the electric seed is $O(s)$ and vanishes at the sphere. This argument concerns the freely specified curl seed. The full initial data also include the conformal and York corrections above.

For the Fourier amplitudes, use smooth scalar potentials $p,U_0,V_0,W_0$ and write
\begin{equation}
 \begin{aligned}
 \varphi&=(1-x^2)p,\qquad W^r=\sqrt{\Delta}(1-x^2)U_0,\\
 W^x&=\sqrt{\Delta}(1-x^2)[-2xV_0+(1-x^2)\partial_x V_0-2iW_0],\\
 W^\phi&=\sqrt{\Delta}[2iV_0-2xW_0+(1-x^2)\partial_x W_0].
 \end{aligned}
\end{equation}
The angular components come from gradient and curl potentials on the sphere, each with the smooth scalar prefactor $(1-x^2)e^{2i\phi}$. Near either bifurcation sphere, the inward radial distance $|r-r_\nu|$ to the horizon $\nu\in\{+,c\}$ is of second order in a signed transverse proper coordinate $\rho$. Choosing $\sqrt{\Delta}\propto \rho$ there makes $W^\rho$ even, the tangential components of $W$ odd and $\varphi$ even. The curl potential $fY_{22}\,\mathrm{d}r$ has an odd normal coefficient in this coordinate. These parities give local smooth extensions of the constructed data across each bifurcation sphere.

We solve for the four potentials as tensor-product polynomials at Gauss--Legendre nodes in $x$ and $z$, where $r=r_++(r_c-r_+)[e^{5(z+1)/2}-1]/[e^5-1]$. The table below reports residuals evaluated by re-differentiating on a separate oversampled grid. The relative volume norms divide the momentum and Hamiltonian residuals by the corresponding constraint source norms, with vector components measured in an orthonormal frame. All three production constraint solves use 18 angular points.

Gauss--Codazzi and the electromagnetic Ricci tensor yield the spacetime Weyl tensor from these spatial fields and derivatives. Its electric part on constrained data is
\begin{equation}
 \begin{aligned}
 \mathcal E_{ij}={}&R^{(3)}_{ij}+KK_{ij}-K_i{}^kK_{kj}+E_iE_j+B_iB_j\\
 &-h_{ij}(E^2+B^2+2\Lambda/3).
 \end{aligned}
\end{equation}
In the normal frame the remaining curvature follows from the Gauss equation, $R^{(4)}_{n_\perp ijk}=D_jK_{ik}-D_kK_{ij}$, and the electrovac Ricci tensor with $R^{(4)}=4\Lambda$, $R^{(4)}_{n_\perp n_\perp}=E^2+B^2-\Lambda$, $R^{(4)}_{n_\perp i}=-2(E\times B)_i$ and $R^{(4)}_{ij}=-2(E_iE_j+B_iB_j)+h_{ij}(E^2+B^2+\Lambda)$. Weyl projection uses the generalized Kinnersley tetrad, with components below ordered as $(t,r,\theta,\phi)$ and $f_\pm=r\pm ia\cos\theta$,
\begin{equation}
 \begin{aligned}
 \ell^\mu&=(r^2+a^2,\Delta,0,a\Xi)/\Delta,\\
 n_{\rm NP}^\mu&=(r^2+a^2,-\Delta,0,a\Xi)/(2\Sigma),\\
 \mathfrak m^\mu&=(ia\sin\theta/T,0,T,i\Xi/(T\sin\theta))/(\sqrt{2}f_+).
 \end{aligned}
\end{equation}
In the orthogonal normal/spatial decomposition we keep the normal-frame components of the tetrad fixed and apply $\delta e_A^i=-h^{ik}\delta h_{kj}e_A^j/2$ to its spatial parts. We use $\Psi_0=C_{\ell\mathfrak m\ell\mathfrak m}$, $\Psi_1=C_{\ell n_{\rm NP}\ell\mathfrak m}$, $\Psi_2=C_{\ell\mathfrak m\bar{\mathfrak m}n_{\rm NP}}$, $\Phi_0=F_{\ell\mathfrak m}$ and $\Phi_1=(F_{\ell n_{\rm NP}}+F_{\bar{\mathfrak m}\mathfrak m})/2$. The background has $\Psi_0=\Psi_1=\Phi_0=0$ and $\Phi_1=Q/(2f_-^2)$. We use $\delta\Psi_0$ and the mixed invariant $\mathcal I_{+1}=2\Phi_1\delta\Psi_1-3\Psi_2\delta\Phi_0$.

Varying the metric requires the symmetric correction above to maintain tetrad normalization. Its one-leg Weyl contractions in $\delta\Psi_0$ vanish on the principal type-D background. In the mixed invariant the separate projections must use a null tetrad for the perturbed metric. Inserting uncorrected background components directly into perturbed tensors can instead give a non-null frame and a different mixed projection. Between normalized tetrads an infinitesimal null rotation about $n_{\rm NP}$ with parameter $b_{\rm rot}$ adds $3b_{\rm rot}\Psi_2$ to $\delta\Psi_1$ and $2b_{\rm rot}\Phi_1$ to $\delta\Phi_0$. These shifts cancel in $\mathcal I_{+1}$. Both $\delta\Psi_0$ and $\mathcal I_{+1}$ are therefore unchanged by the Maxwell-dependent rotation used below.

With $\nu_0=\delta\Phi_0/(2\Phi_1)$, a first-order null rotation about $n_{\rm NP}$ with parameter $-\nu_0$ sets the positive extreme Maxwell scalar to zero and gives \cite{WellerEtAl2026Parity}
\begin{equation}
 \begin{aligned}
 \delta\Psi_{1,\mathrm{ph}}&=\mathcal I_{+1}/(2\Phi_1),\\
 \kappa_{\mathrm{ph}}&=\delta\kappa+(D_{\ell}-\rho_{\rm NP})\nu_0,\\
 \sigma_{\mathrm{ph}}&=\delta\sigma+(\delta_{\mathfrak m}-2\beta_{\rm NP}-\tau_{\rm NP})\nu_0.
 \end{aligned}
\end{equation}
Here $D_\ell=\ell^\mu\partial_\mu$ and $\delta_{\mathfrak m}=\mathfrak m^\mu\partial_\mu$ are background directional derivatives, and $\delta\kappa,\delta\sigma$ are the first variations of $-g(\mathfrak m,\nabla_{\ell}\ell)$ and $-g(\mathfrak m,\nabla_{\mathfrak m}\ell)$ in the symmetric tetrad. A centered projection check on the central data at three exterior points and two Fourier phases gives absolute changes below $3.8\times10^{-28}$ in $\delta\Psi_0$ and $1.3\times10^{-30}$ in $\mathcal I_{+1}$ under this rotation at a difference step of $10^{-12}$. The rotation changes the individual Maxwell and Weyl projections and the connection coefficients $\kappa,\sigma$ used in the canonical map. The subsequent rotation setting $\delta\Phi_2=0$ leaves the positive-spin fields in this map unchanged to first order. To obtain the time derivatives in the optical map we hold the background lapse $N$ and shift $\beta$ fixed on the slice and vary the equations
\begin{equation}
 \begin{aligned}
 (\partial_t-\mathcal L_\beta)h_{ij}&=-2NK_{ij},\\
 (\partial_t-\mathcal L_\beta)E^i&=NK E^i+\varepsilon^{ijk}D_j(NB_k),\\
 (\partial_t-\mathcal L_\beta)B^i&=NK B^i-\varepsilon^{ijk}D_j(NE_k).
 \end{aligned}
\end{equation}
Here $K=\operatorname{tr}_h K$. The resulting velocities are determined by the Cauchy data and their spatial derivatives.

For the canonical map define $B_0=(\delta\Psi_0,-2Q\kappa_{\mathrm{ph}}/(\sqrt{2}f_-^2))^{\mathsf T}$, $B_1=(\sqrt{2}f_-\delta\Psi_{1,\mathrm{ph}},-2Q f_+\sigma_{\mathrm{ph}}/f_-^2)^{\mathsf T}$ and $\mathcal P=\Delta^{5/4}\sqrt{T\sin\theta}$. Let $X$ be Zhao's frequency-independent matrix in the $\varsigma=+1$ transformation \cite{Zhao2026KNAdS}. The two chiral fields and the initial canonical vector $q_0=q(0)$ are
\begin{equation}
 \begin{aligned}
 y_\pm&=\mathcal P(B_0\pm i B_1/\sqrt\Delta),\\
 z_+&=\operatorname{adj}(X)y_+,\qquad z_-=X^\dagger y_-,\\
 v&=z_++z_-,\qquad w=-i\sigma_2(z_+-z_-),\\
 q_0&=(v,w)^{\mathsf T}.
 \end{aligned}
\end{equation}
In the implementation the two combinations are simplified algebraically before evaluation. With $X=X_0+\sqrt\Delta X_1$, one uses $X_0^\dagger=-\operatorname{adj}X_0$ and $X_1^\dagger=\operatorname{adj}X_1$ on the real exterior. This gives
\begin{equation}
 \begin{aligned}
 \Delta^{-3/4}v&=2\sqrt{T\sin\theta}\,[i\operatorname{adj}(X_0)B_1+\operatorname{adj}(X_1)\Delta B_0],\\
 \Delta^{-1/4}w&=-2i\sqrt{T\sin\theta}\,\sigma_2\bigl[\operatorname{adj}(X_0)\Delta B_0\\
 &\qquad+i\Delta\operatorname{adj}(X_1)B_1\bigr].
 \end{aligned}
\end{equation}

The full fixed-$m$ frequency-domain operator on $q=(v,w)^{\mathsf T}$ is
\begin{equation}
 \begin{aligned}
 \mathcal{Q}(\omega)&=\partial_r-\frac{R_0\sigma_3+R_1\sigma_1+h_r\sigma_2}{\sqrt\Delta}\otimes I_2\\
 &\quad-\frac{\sigma_2}{\sqrt\Delta}\otimes\mathcal A_\theta(\omega),\\
 \mathcal A_\theta(\omega)&=-iT\sigma_2\partial_\theta-\Theta_1\sigma_3+\Theta_0\sigma_1+h_\theta I_2 .
 \end{aligned}
\end{equation}
The time equation is $(\mathcal{Q}_0+i\mathcal{Q}_\omega\partial_t)q=0$. For $\widetilde q=\int_0^\infty e^{i\omega t}q(t)\,\mathrm{d}t$, integration by parts gives $\mathcal{Q}(\omega)\widetilde q=i\mathcal{Q}_\omega q_0$. The pointwise time matrix satisfies
\begin{equation}
 \begin{aligned}
 \mathcal{Q}_\omega^2&=-c_{\rm time}I_4,\\
 c_{\rm time}&=\frac1\Delta\left[\frac{(r^2+a^2)^2}{\Delta}-\frac{a^2\sin^2\theta}{T^2}\right]>0.
 \end{aligned}
\end{equation}
Let $\mathcal{Q}_\theta=iT(\sigma_2\otimes\sigma_2)/\sqrt\Delta$ be the coefficient of $\partial_\theta$. The Pauli algebra gives $\{\mathcal{Q}_\omega,\mathcal{Q}_\theta\}=0$ and $\mathcal{Q}_\theta^2=-(T^2/\Delta)I_4$. For $H_\xi=-i\mathcal{Q}_\omega^{-1}(\xi_rI_4+\xi_\theta \mathcal{Q}_\theta)$, this yields
\begin{equation}
 \begin{aligned}
 H_\xi^2&=v_\xi^2I_4,\qquad v_\xi^2=\frac{\xi_r^2+T^2\xi_\theta^2/\Delta}{c_{\rm time}},\\
 S_\xi&=\Pi_+^\dagger \Pi_++\Pi_-^\dagger \Pi_->0,\\
 \Pi_\pm&=\tfrac12(I_4\pm H_\xi/v_\xi).
 \end{aligned}
\end{equation}
For nonzero covectors on compact exterior coordinate patches, $S_\xi H_\xi=H_\xi^\dagger S_\xi$. This gives local strong hyperbolicity and Cauchy uniqueness. The fixed-$m$ evolution uses the regular spin-weighted axis domain, implemented by the weights of Appendix~\ref{app:evolution-detuning}. A constrained Einstein--Maxwell development provides a solution of the radiative equations with the same Cauchy map. The comparison therefore uses a forward radiative evolution of physical initial data. The principal-symbol argument here is local to the smooth exterior and supplies no global estimate at the horizon or coordinate axes.

Independent tensor and compiled curvature maps agree to $7.2\times10^{-16}$ relative at sampled interior points; NH checks give $6.2\times10^{-11}$ and $2.2\times10^{-12}$. A direct null-rotated connection calculation checks the optical map to $1.6\times10^{-10}$ absolute on its grid. The time-matrix algebraic identity passes to $2.6\times10^{-15}$. These checks are distinct from physical-source refinement.

\begin{table}[tbp]
\caption{Oversampled differential constraint residuals for the three sources, measured in relative volume $L^2$ norm.}
\centering
\begin{tabular}{lrrr}\hline\hline
Source & Radial nodes & Momentum & Hamiltonian\\\hline
Inner &128&$9.4\times10^{-8}$&$5.2\times10^{-11}$\\
Central&160&$1.8\times10^{-8}$&$8.4\times10^{-13}$\\
Outer&128&$3.1\times10^{-9}$&$3.4\times10^{-11}$\\\hline\hline
\end{tabular}
\end{table}

For the finite-radius extraction in the Letter, a local Carter frame rescales $\mathcal O$ by $\Delta/(2\Sigma)\simeq0.221$ at the observer. The proper-time frequency is $u^t(\omega-m\Omega_{\rm Car})$, with $u^t=\mathrm{d}t/\mathrm{d}\tau$, $\tau$ the observer's proper time, and $\Omega_{\rm Car}=a\Xi/(r^2+a^2)$. Both transformations preserve the pole order.

\section{Physical Green functional and endpoint continuation}
\label{app:green-functional}
For the selected simple angular branch solve $\mathcal A_\theta(\omega)\chi=\lambda(\omega)\chi$, with $\chi_2(\pi/2)=1$. Set $\mathbf s=i\mathcal{Q}_\omega q_0$ and denote its two angular blocks by $\mathbf s_1,\mathbf s_2$. The projected radial forcing $\mathbf j$ has components
\begin{equation}
 (\mathbf j)_k(r,\omega)=\frac{\displaystyle\int_0^\pi\chi(\theta,\omega)^{\mathsf T}\mathbf s_k(r,\theta)\,\mathrm{d}\theta/T}{\displaystyle\int_0^\pi\chi(\theta,\omega)^{\mathsf T}\chi(\theta,\omega)\,\mathrm{d}\theta/T},\qquad k=1,2.
\end{equation}
The projection is bilinear, with the holomorphic eigenfunction and its norm recomputed at each frequency. For radial amplitudes $u$ and observer angle $\theta_o=\pi/2$, the extraction uses $v=u_1\chi$, $w=u_2\chi$ and $z_\pm=(v\pm i\sigma_2 w)/2$. Invert the transformation above by $y_+=Xz_+/\det X$ and $y_-=(X^\dagger)^{-1}z_-$. The curvature functional is $\mathcal L_0 u=(y_{+,1}+y_{-,1})/(2\mathcal P)$, evaluated at $r=3,\theta_o$. For the mixed invariant use $(Q/(\sqrt{2}f_-^3))\sqrt\Delta(y_{+,1}-y_{-,1})/(2i\mathcal P)$ in the same frame.

Let $E,C$ now denote event-ingoing and cosmological-outgoing radial solutions with $E_2(3)=C_2(3)=1$. In this chart $D=E_1(3)-C_1(3)=\det(E,C)$. With $J=i\sigma_2$, the projected forcing $\mathbf j$ gives
\begin{equation}
 \begin{aligned}
 N_{\rm phys}&=\mathcal L_0[C(3)I_E+E(3)I_C],\\
 I_E&=\int_{r_+}^{3}E^{\mathsf T}J\mathbf j\,\mathrm{d}r,\qquad I_C=\int_3^{r_c}C^{\mathsf T}J\mathbf j\,\mathrm{d}r.
 \end{aligned}
\end{equation}
All propagation rescalings and Frobenius normalizations are restored. The selected angular contribution is $\widetilde{\mathcal O}_{\rm br}=N_{\rm phys}/D$. For an isolated double zero of this branch,
\begin{equation}
 \begin{aligned}
 C_{-2}&=\frac{2N_{\rm phys}(\omega_{\rm EP})}{D_{\omega\omega}(\omega_{\rm EP})},\\
 C_{-1}&=\frac{2N_{{\rm phys},\omega}}{D_{\omega\omega}}-\frac{2N_{\rm phys}D_{\omega\omega\omega}}{3D_{\omega\omega}^2}.
 \end{aligned}
\end{equation}
All quantities on the right-hand sides are evaluated at $\omega_{\rm EP}$. The frequency derivatives follow the selected angular branch and include the angular projection. We use a homogeneous scattering transfer in the same radial chart and extraction convention as a reference, with $N_{\rm ref}(\omega_{\rm EP})\ne0$. Set $S=(I_E+I_C)\mathcal L_0 E(3)/N_{\rm ref}$. Because $C(3)-E(3)=(-D,0)^{\mathsf T}$, this equals $N_{\rm phys}/N_{\rm ref}$ up to a term proportional to $D$. Thus $C_{-2}=SC_{-2}^{\rm ref}$ and $C_{-1}=SC_{-1}^{\rm ref}+S_\omega C_{-2}^{\rm ref}$ at $\omega_{\rm EP}$. We compute $S_\omega$ by centered frequency differences with Richardson extrapolation. For the central data the two step pairs $(2.5,5)\times10^{-6}$ and $(5,10)\times10^{-6}$ change $C_{-1}$ by $7.7\times10^{-8}$ relative. At a simple pole $\omega_n$, the residue for initial profile $j$ is $R_{nj}=N_{{\rm phys},j}(\omega_n)/D_\omega(\omega_n)$, where $N_{{\rm phys},j}$ denotes the numerator for that profile. The complete time evolution in Appendix~\ref{app:evolution-detuning} retains the other angular components.

The overlap integrals are defined first where they converge in the retarded upper half plane. For the horizon index $\nu\in\{+,c\}$, use $x_+=r-r_+$ and $x_c=r_c-r$. Set $\varpi_\nu=(\omega-m\Omega_\nu)/(2\kappa_\nu)$ and $P_\nu=\Delta/x_\nu$. The weighted forcing components are $H_1=\Delta^{1/4}(\mathbf j)_1$ and $H_2=\Delta^{3/4}(\mathbf j)_2$. The powers in Appendix~\ref{app:geometry} then give the two integrands as $x_\nu^{\alpha_\nu}g_\nu(x_\nu)$ with
\begin{equation}
 \begin{aligned}
 \alpha_+&=-1-i\varpi_+,\quad & g_+&=a_+P_+^{-3/4}H_2-b_+P_+^{-1/4}H_1,\\
 \alpha_c&=-i\varpi_c,\quad & g_c&=a_cP_c^{-3/4}H_2-x_cb_cP_c^{-1/4}H_1.
 \end{aligned}
\end{equation}
Here $a_\nu,b_\nu$ are the regular Frobenius amplitudes of the first and second radial components after extracting their leading powers of $x_\nu$. They absorb the regular phase and the regular factors from $\Delta=x_\nu P_\nu$ in the homogeneous radial solutions. Changing variable in the cosmological integral reverses its bounds and contributes a positive integral from $x_c=0$ to the matching point. At the EP $\alpha_+\simeq-3.69243-0.87041i$ gives a divergent real event overlap, while $\alpha_c\simeq-0.63156-16.33886i$ gives an integrable and rapidly log-oscillatory cosmological endpoint. The resonant weight causes the event-overlap divergence after continuation even for the regular curl seed described in Appendix~\ref{app:radiative-map}.

For a local analytic factor $g(x^{1/p})$, write $a_n=c_n\rho^{n/p}$ for its Taylor coefficients $c_n$ scaled to the circle $|x|=\rho$. The continuation is
\begin{equation}
 \mathfrak I(\alpha;\rho)=\frac{\int_{\Gamma_{\rho,p}}x^\alpha g(x^{1/p})\,\mathrm{d}x}{e^{2\pi ip(\alpha+1)}-1}=\rho^{\alpha+1}\sum_{n\ge0}\frac{a_n}{\alpha+1+n/p},
\end{equation}
where $\Gamma_{\rho,p}$ follows $p$ counterclockwise circuits from the positive real axis with continuously tracked logarithm and roots. Termwise integration where the endpoint converges gives the identity. The series distinguishes removable zeros of the monodromy denominator from genuine poles. In the computation, after analytic quarter-weighting, both reduced factors use $p=1$. We sample the periodic analytic factor alone and take a discrete Fourier transform to estimate the $a_n$. The negative-frequency half is monitored for aliasing and excluded from the Taylor sum. The typical radii are $(\rho_+,\rho_c)=(0.003,0.1)$ with 128 nodes each and 20 angular quadrature points. The event circle lies inside the inner-horizon separation. At the central EP with source-160 data the discarded Fourier mode levels relative to the sampled factors are $2.6\times10^{-13}$ and $1.1\times10^{-9}$ at the event and cosmological endpoints, respectively. The two local contributions are added to bulk Gauss quadratures in the same $D$ chart.

Independent checks with the fixed source-96 data give event-local circle versus direct real-axis overlap agreement of $3.6\times10^{-13}$ at the upper-plane test frequency $\omega_p=\operatorname{Re}\omega_{\rm EP}+0.04i$ and radius $0.003$. The event-local radius difference $0.003\to0.004$ agrees with direct finite-annulus integration at the EP to $3.6\times10^{-12}$. For the cosmological endpoint, independent real-axis Chebyshev interpolation with exact integration of the Mellin powers agrees at radius $0.1$ to $1.5\times10^{-8}$ in the upper plane and $2.5\times10^{-8}$ at the EP. The complete selected-branch continued Green transfer agrees with the independent direct upper-plane calculation to $4.3\times10^{-11}$ for curvature and $5.8\times10^{-11}$ for the mixed invariant. These are fixed-data checks, separate from constraint-solution refinement and time-evolution closure.

As a separate event-endpoint check, we sample the analytic factor at 10, 14 and 18 Chebyshev nodes on the real exterior and continue its polynomial moments. We hold the initial slice, radial modes, bulk quadrature and cosmological overlap fixed. Replacing the event circle changes $C_{-2}$ by less than $3.4\times10^{-6}$ relative for all three production profiles. Recomputing $C_{-1}$ and all seven NH residues for the central profile shifts the first crossover of the coherent NH modulus and the EP-pair modulus by less than $0.0014M$ from $223.905M$. The 18-node check amplifies roundoff in the highest NH residue, while the 10- and 14-node checks give smaller changes. The archive contains the three replays.

The central $128\to160$ source refinement changes $C_{-2}$ by $9.36\times10^{-7}$ relative; earlier $64\to96$ and $96\to128$ changes are $5.12\times10^{-5}$ and $1.10\times10^{-5}$. At fixed source, circle radii/nodes, angular order and bulk quadrature change it by less than $5\times10^{-13}$. Inner $96\to128$ and outer $96\to128$ changes are $5.96\times10^{-5}$ and $4.99\times10^{-7}$. Numerical arithmetic precision is therefore not the physical-source error scale. Higher-NH residues change by $4.6\times10^{-6}$ to $1.7\times10^{-5}$ under central source refinement; changed-setting root defects are below $6\times10^{-14}$.

Using the production central source numerator and the independent multiprecision value of $D_{\omega\omega}$ gives $C_{-2}=(-7.1804511823-8.8076768929i)\times10^{-7}$, agreeing with the scattering-contour calibration to $1.52\times10^{-12}$ relative. The source refinement sets the larger numerical scale. Recomputing $C_{-1}$ at fixed radial and angular kernel settings changes the central curvature coefficient by $6.65\times10^{-7}$ between the 128- and 160-point source solves. The older 96-point result differs from the 160-point value by $6.96\times10^{-6}$. The direct Laurent expression for $C_{-1}$ also reproduces the quoted $A$ to its displayed precision.

The squared electric-seed norms are $N_e^2\simeq4.78688,12.17175,23.12410$ for the inner, central and outer data. These norms refer to the freely specified electric seed. The common sign of the normalized $C_{-2}$ values extends to nonnegative real mixtures by linearity. For the tuned combination, $w_{\rm mix}=-B_{\rm central}/B_{\rm inner}$ with the central-160 and inner-128 coefficients. The earlier inner-96 coefficient gives a residual scale of $5\times10^{-11}$ for the tuned $C_{-2}$, which gives an empirical scale for the cancellation.

\begin{figure}[t]
	\centering
	\includegraphics[width=\columnwidth]{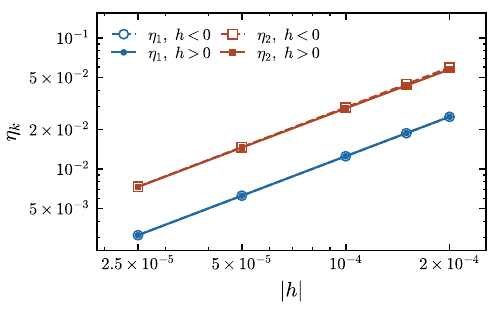}
	\caption{Convergence of the residue moments for the central electric seed, evaluated in the curvature channel. For $h=a-a_{\rm EP}$ at fixed $\Lambda$ and $\epsilon=1-r_-/r_+$, the split-pole residues $R_\pm$ give $C_{-1}(h)=R_++R_-$ and $C_{-2}(h)=\delta(R_+-R_-)$, with $\delta=(\omega_+-\omega_-)/2$. The plotted quantity is $\eta_k=|C_{-k}(h)-C_{-k}(0)|/|C_{-k}(0)|$, where $k\in\{1,2\}$ labels the Laurent order and $C_{-k}(0)$ is the corresponding EP moment. Each initial-data solve uses 96 radial points. Markers denote sampled detunings and line segments connect successive samples of the same sign.}
	\label{fig:residue-moments}
\end{figure}

For each source $j$ define $\widehat F_j=F_j/\|F_j\|$ and $\widehat P_j=P_j/\|F_j\|$ using the same evolved norm. For $u=e^{0.572844i}\widehat F_{\rm central}$, the real projections $r_j=\operatorname{Re}(u^\dagger\widehat F_j)$ are $(0.840363,0.651847,0.979641)$ for central, inner and outer signals on $250$--$300M$. With $e_j=\|\widehat F_j-\widehat P_j\|$, the triangle inequality bounds the residual by $\sum_j\alpha_j e_j$ and projection on $u$ bounds the denominator below by $\sum_j\alpha_jr_j$. Thus for nonzero nonnegative mixtures,
\begin{equation}
 \frac{\|\sum_j\alpha_j(\widehat F_j-\widehat P_j)\|}{\|\sum_j\alpha_j\widehat F_j\|}
 \leq\max_j\frac{e_j}{r_j}<0.00116,\qquad \alpha_j\geq0 .
\end{equation}
The bound applies to the computed sampled sequences.

\section{Evolution and detuning controls}
\label{app:evolution-detuning}
The full fixed-$m$ radiative evolution removes exact weights using $v=\Delta^{3/4}\mathsf P_{\rm ax}(x)V$ and $w=\Delta^{1/4}\mathsf P_{\rm ax}(x)W$, with $x=\cos\theta$ and $\mathsf P_{\rm ax}(x)=\operatorname{diag}[(1-x)^{7/4}(1+x)^{3/4},(1-x)^{9/4}(1+x)^{1/4}]$. Use $r=(r_++r_c)/2+(r_c-r_+)\tanh(ku)/(2\tanh k)$ with $k=8$ and uniform cell centers in $-1<u<1$. Five-point fourth-order radial derivative stencils are centered in the interior and shifted at the two ends. Angular differentiation uses polynomial collocation at Gauss--Legendre nodes after factoring the axis weights. Extraction interpolates in angle and uses eight neighboring radial nodes. The time integrator is DOP853 with absolute tolerance $10^{-3}$ times its relative tolerance. Sixth-order radial Kreiss--Oliger dissipation uses coefficient $0.02$ for production. The finest central run is $1536\times18$ with relative tolerance $2\times10^{-10}$; the inner and outer runs use $1536\times14$ and $2\times10^{-9}$.

With $\mathrm{d}r_*/\mathrm{d}r=(r^2+a^2)/\Delta$, the principal cone in Appendix~\ref{app:radiative-map} gives $|\mathrm{d}r_*/\mathrm{d}t|<1.05$ throughout the central exterior. The observer-centered endpoint tortoise distances are approximately $-1456M$ and $607M$ on the finest grid. They exclude continuum causal contact between the cutoffs and the extraction through $300M$. Grid and map refinement test the numerical closure separately. The distance from the observer to $r=r_++10^{-4}M$ is $561.5M$. The same cone bound therefore excludes influence through $500M$ from alterations confined to $0<r-r_+\leq10^{-4}M$ when the exterior Cauchy data elsewhere are held fixed.

\begin{table}[tbp]
\caption{Central full-curvature convergence controls. Relative complex $L^2$ changes on $250\leq t/M\leq300$, except the last two on $200\leq t/M\leq250$. Angular/map comparisons also retain small time-tolerance changes; the combined row is labeled explicitly.}
\centering
\begin{tabular}{@{}l@{\hspace{2pt}}r@{}}\hline\hline
Change & Relative change\\\hline
Radial $768\to1024$ at 14 angular points&$7.36\times10^{-6}$\\
Combined $1024\times14\to1536\times18$&$2.91\times10^{-6}$\\
Angular $14\to18$ at 1536 radial points&$1.14\times10^{-6}$\\
Map $k=7.5\to8$&$3.81\times10^{-6}$\\
Physical data $96\to160$&$5.01\times10^{-7}$\\
Time tolerance refinement&$3.94\times10^{-6}$\\
Dissipation $0.02\to0.01$&$6.84\times10^{-7}$\\\hline\hline
\end{tabular}
\end{table}

An auxiliary ODE accumulates the upper-plane transform at $\omega_p=\operatorname{Re}\omega_{\rm EP}+0.04i$. This diagnostic projects the time-evolved fields onto the same angular branch used by the Green calculation. The two calculations share the physical initial-data map. Curvature discrepancies are $4.00\times10^{-4}$, $1.21\times10^{-4}$ and $2.30\times10^{-5}$ at 768, 1024 and 1536 radial points, consistent with fourth-order convergence. The mixed invariant gives $2.20\times10^{-4}$, $7.17\times10^{-5}$ and $1.46\times10^{-5}$.

For each profile, both Laurent coefficients, all seven NH residues and the corresponding evolution use the same constraint solution, with 160 radial points for the central source and 128 for the inner and outer sources. Earlier mixed-resolution arrays in the archive document convergence. The displayed waveform comparisons use the matched records. For the inner and outer data, recomputing $C_{-1}$ and the NH residues $R_{nj}$ with $n=0,\ldots,3$ for each source $j$ at 128 radial source points changes $C_{-1}$ by $3.56\times10^{-5}$ and $2.87\times10^{-7}$, respectively. The largest lower-NH residue changes in the curvature channel are $3.48\times10^{-5}$ and $3.15\times10^{-7}$. Halving the Richardson frequency steps changes $C_{-1}$ by $1.28\times10^{-6}$ and $1.17\times10^{-6}$. Refining the time-evolution radial grid from 1024 to 1536 changes the two late curvature vectors by $5.94\times10^{-6}$ and $6.33\times10^{-6}$. The tuned prediction combines the matched pole predictions with the same complex weight used for the time signals.

\begin{table}[tbp]
\caption{Central curvature discrepancies in percent for the four consecutive windows highlighted in Fig.~\ref{fig:waveform}. Ordinary fits retain all seven calculated NH terms. For the free-frequency model, the frequency is fitted separately in each window.}
\label{tab:central-fits}
\centering
\begin{tabular}{rrrrr}\hline\hline
\shortstack{Window\\$t/M$} & \shortstack{Physical\\Jordan+NH} & \shortstack{Fit $A$,\\fixed $\omega_{\rm EP}$} & Fit $A,\omega$ & \shortstack{Damping\\bias (\%)}\\\hline
100--150&25.44&23.29&19.4&$+47.3$\\
150--200&5.368&5.489&4.619&$-8.32$\\
200--250&0.6049&1.583&0.362&$-3.22$\\
250--300&0.09350&0.9610&0.0405&$-2.73$\\\hline\hline
\end{tabular}
\end{table}

On the 26 central curvature samples from $250M$ to $300M$, the norms of the seven-NH sum and the EP pair are $0.6734$ and $0.4022$ times the evolved norm, respectively. Their phases enter the total discrepancy through interference. For this source the EP damping time is $45.04M$, and the ratio $|Bt|/|A|$ grows from $0.08358$ at one damping time to unity at $538.9M$. The curvature rms on $250$--$300M$ is $2.15\times10^{-4}$ of the pulse peak. Deleting only $Bt$ gives a discrepancy of $18.7769\%$. At fixed $\omega_{\rm EP}$ and fixed NH terms, fitting $A$ with $B=0$ gives $0.9610\%$, while fitting both complex coefficients gives $0.03475\%$. The latter fit gives $|B_{\rm fit}-B|/|B|=0.0242$, conditional on the fixed frequency and NH prediction.

We also fit a single free complex frequency and amplitude to the prescribed central Jordan-plus-NH prediction itself, keeping the seven NH terms fixed on the same samples. The fit gives $M\omega=0.57624756-0.02159284i$ with discrepancy $0.01119\%$ and damping bias $-2.7435\%$. The fit to the evolved waveform gives $M\omega=0.57629217-0.02159580i$ and damping bias $-2.7302\%$. Both fits eliminate the amplitude by complex least squares at each frequency. In the control with $B=0$ in the prescribed waveform, the fit recovers $\omega_{\rm EP}$ to numerical precision. For the prescribed pair alone, $i\,\mathrm{d}\ln[(A+Bt)e^{-i\omega_{\rm EP}t}]/\mathrm{d}t=\omega_{\rm EP}+iB/(A+Bt)$. The real part of the shift decreases from $1.689\times10^{-3}/M$ to $1.587\times10^{-3}/M$ across the window. The four full complex pole separations in Fig.~\ref{fig:unfolding} range from $4.594\times10^{-4}/M$ to $9.310\times10^{-4}/M$.

At each detuning, the constraints, angular projections, nine frequencies, source residues and both endpoint continuations are recomputed. The time evolution uses $1024\times14$ points and source-96 data. Radial $768\to1024$ changes the late curvature by at most $7.4\times10^{-6}$. Upper-plane time/frequency closure is $1.21\times10^{-4}$ for curvature and $7.17\times10^{-5}$ for the mixed invariant. Changed-setting frequency defects are at most $1.6\times10^{-11}$. At $h=\pm2\times10^{-4}$, source-96 to source-128 changes residues by less than $3.9\times10^{-5}$ and the coherent nine-pole late signal by at most $5.3\times10^{-8}$. Doubling both source-circle node counts at $h=+2\times10^{-4}$ with source-128 data changes residues by less than $3.9\times10^{-12}$.

\begin{table}[H]
\caption{Detuned full-curvature discrepancies in percent on $250\leq t/M\leq300$. The local Jordan alternative uses the detuned midpoint and moments with $\cos\delta t\to1$ and $\sin(\delta t)/\delta\to t$. The frozen alternative keeps the central EP pair. All retain the seven detuned NH terms.}
\label{tab:detuned-waveforms}
\centering
\begin{tabular}{rrrr}\hline\hline
$h$ & \shortstack{Split pair\\+NH} & \shortstack{Local Jordan\\+NH} & \shortstack{Frozen EP\\+NH}\\\hline
$-2\times10^{-4}$&0.09411&0.36819&1.68602\\
$-5\times10^{-5}$&0.09384&0.15190&0.47437\\
$+5\times10^{-5}$&0.09353&0.07130&0.32120\\
$+2\times10^{-4}$&0.09286&0.22811&1.46579\\\hline\hline
\end{tabular}
\end{table}
At $h=+5\times10^{-5}$, let $e=F-P_{\rm split}$ and $d=P_{\rm local}-P_{\rm split}$ on the late samples. In the squared local-model residual, the term $\|d\|^2/\|F\|^2=5.01\times10^{-7}$ is outweighed by the cross term $-2\operatorname{Re}(e^\dagger d)/\|F\|^2=-8.67\times10^{-7}$. The local truncation error therefore partially cancels the existing residual on this window.

\section{Extended waveform comparison}
\label{app:extended-waveforms}

We evolve the central source-160 initial data to $500M$ on $768\times14$, $1024\times14$ and $1536\times18$ radial--angular grids with the mapping and dissipation of Appendix~\ref{app:evolution-detuning}. The DOP853 relative and absolute tolerances are $2\times10^{-10}$ and $2\times10^{-13}$. The shortest observer-to-cutoff tortoise distance among these grids is $587M$, which exceeds $1.05\times500M$. The characteristic bound in Appendix~\ref{app:evolution-detuning} therefore excludes continuum causal contact with either cutoff during the extended runs.

For each observable we compare the calculated double-pole term with an ordinary exponential of complex amplitude and frequency. The two models retain the same seven calculated NH terms. At each trial frequency we eliminate the complex amplitude by least squares and use 21 starts for the bounded frequency search. The late refit varies both exponential parameters on the 76 samples $t/M=350,352,\ldots,500$. The forecast holds the exponential parameters from the archived $250$--$300M$ samples fixed and evaluates them on the later samples. The Jordan prediction uses the calculated source coefficients in both intervals. Table~\ref{tab:extended-ivp} shows the sampled relative complex norms in percent. The two finest evolutions differ by $0.000522\%$ in curvature and $0.000455\%$ in the mixed invariant on this interval, using the finer trace for normalization.

\begin{table*}[tp]
\caption{Extended-window discrepancies in percent for the central source on $350\leq t/M\leq500$. The ordinary exponential is either refitted on these samples or forecast from the fit on $250\leq t/M\leq300$. Fits are performed separately for the two observables. All three models retain the calculated NH terms.}
\label{tab:extended-ivp}
\centering
\begin{tabular}{@{}crrrrrr@{}}
\hline\hline
 & \multicolumn{3}{c}{Curvature} & \multicolumn{3}{c}{Mixed invariant}\\
\cline{2-4}\cline{5-7}
Grid & Jordan & Refit & Forecast & Jordan & Refit & Forecast\\
\hline
$768\times14$  & 0.00502 & 0.01393 & 0.26163 & 0.00491 & 0.01382 & 0.22425\\
$1024\times14$ & 0.00413 & 0.01392 & 0.26196 & 0.00398 & 0.01383 & 0.22451\\
$1536\times18$ & 0.00383 & 0.01401 & 0.26207 & 0.00369 & 0.01388 & 0.22460\\
\hline\hline
\end{tabular}
\end{table*}

On $350$--$500M$ the norms of the coherent NH sum and the EP pair are $0.8824$ and $0.1305$ times the evolved curvature norm, respectively. The curvature rms is $4.78\times10^{-5}$ of the sampled pulse peak. Fitting the complex coefficients $A$ and $B$ at fixed $\omega_{\rm EP}$ and fixed NH terms on the finest grid recovers the calculated $B$ within $0.22\%$ in both observables. On the full held-out interval $302$--$500M$, the physical and forecast curvature discrepancies are $0.01518\%$ and $0.18418\%$, respectively.

As a sensitivity test we allow one complex scale $\alpha_{\rm NH}$ on the coherent NH sum. On the finest grid and $350$--$500M$, the ordinary fit with complex $A,\omega,\alpha_{\rm NH}$ gives curvature and mixed-invariant discrepancies of $0.003223\%$ and $0.003247\%$. The fit with complex $A,B,\alpha_{\rm NH}$ and fixed $\omega_{\rm EP}$ gives $0.000317\%$ and $0.000253\%$, respectively. These fits have the same parameter count and keep the relative NH coefficients fixed.

The endpoint checks and the waveform evolution use the same stationary initial slice. A comparison with a horizon-penetrating evolution would require transporting that Cauchy solution to the alternative slice. The numerical reproduction archive accompanying this submission contains the waveforms, coefficients, fits and replay commands.

\bibliography{biblio}

\end{document}